\documentclass[12pt,a4paper]{article}

\usepackage{graphics}
\usepackage{latexsym}
\usepackage{amsmath}
\usepackage{amssymb}

\title{A model of memory, learning and recognition}

\author{Bruce Hoeneisen}
\date{\small{Universidad San Francisco de Quito \\
        6 May 2002}}

\begin{document}
\maketitle

\begin{abstract}
\noindent
We propose a simple model of recognition,
short-term memory, long-term memory and learning.
\end{abstract}

\section{Introduction}

For years I have been thinking
about how conscience emerges from
the underlying neuron clatter (by conscience I
mean awareness, \textit{i.e.}
the difference between being awake and asleep).
Last night the inspiration came all at once, just
as visitors were arriving. I put my thoughts aside
due to the visitors, but today I hasten to write them
down. The model explains a lot and in
hindsight is almost \textquotedblleft{obvious}" (why did I not think
of it before?) I mention these details because they
illustrate how the mind works.

For the sake of simplicity I will be very specific,
almost telegraphic. I will mention only the essential
points. The reader should not take seriously the
specific details. I have no pretense that the brain works with
exactly the implementation I will describe below.
However I do believe that many of the ideas discussed
here are in fact implemented by the brain in some
equivalent way.

To make this article self-contained I list
in Section 2 characteristics of the brain
that are relevant to our model.[1-6]
Some of these
characteristics have been obtained directly
from experiments or observations and some have
been inferred indirectly and are therefore
only educated guesses. Our model is presented
in Section 3. Estimates of the number of memory
neurons are obtained in Section 4.
Finally, conclusions are collected in Section 5.

\begin{figure}
\begin{center}
\scalebox{0.6}
{\includegraphics{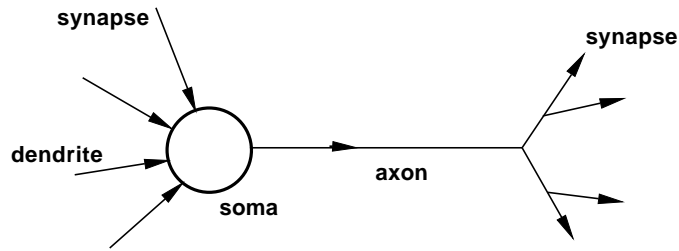}}
\caption{Schematic drawing of a neuron.}
\label{neuron}
\end{center}
\end{figure}

\section{Characteristics of the brain}

The human brain has approximately $10^{10}$ \textquotedblleft{neurons}".
A neuron is a cell with a central body (the \textquotedblleft{soma}")
and up to $\approx10^5$ ramifications. One of these
ramifications, called \textquotedblleft{axon}", is long and carries
the output electrical pulse of the neuron.
These pulses have an amplitude of 145mV and a
duration of about 0.2 milliseconds.
The remaining ramifications, called \textquotedblleft{dendrites}",
carry the input electrical pulses to the neuron.
The axon is connected to
dendrites of other neurons (or to muscle
cells). These connections, or \textquotedblleft{synapses}", are
of a chemical nature. Each synapse carries a
\textquotedblleft{weight}" that can be excitatory or inhibitory.
The neuron \textquotedblleft{fires}",
\textit{i.e.} sends an output impulse along its axon, if
the weighted sum of inputs is above a threshold.
A schematic diagram of a neuron is shown in
Figure \ref{neuron}.

\textquotedblleft{There are roughly a quarter of a million nerve
cells and two billion synapses below one square
millimeter of cerebral-cortex surface.}"\cite{Koch}

The programing of the brain consists of the
making or breaking of synapses (or more generally
of the modification of the weights of the synapses).
Certain programing occurs at critical periods
(such as the breaking of synapses to obtain
registration of the images seen by the right
and left eyes\cite{Delbruck}) and programing that occurs
throughout the lifetime (such as storing
information in the long-term memory).

Many brain processes occur at the same time:
parallel computation is widespread.

\begin{figure}
\begin{center}
\scalebox{0.6}
{\includegraphics{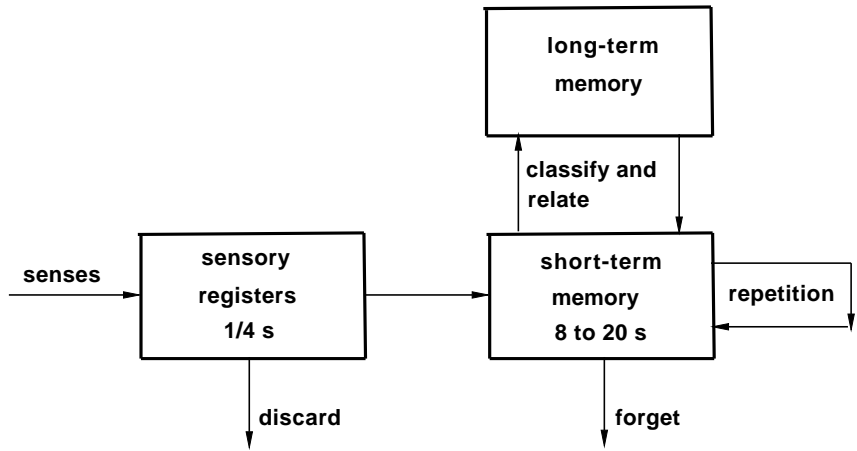}}
\caption{Memory as seen by a psychologist.}
\label{memory}
\end{center}
\end{figure}

Let us describe \textquotedblleft{sensory registers}", \textquotedblleft{short-term
memory}" and \textquotedblleft{long-term memory}" from the point
of view of a psychologist. A block diagram
is shown in Figure \ref{memory}.
The stimuli from the environment arrive at our
sense organs where they are transformed into
electro-chemical impulses in neurons. This
information is processed and analyzed
hierarchically in the
sensory registers, and in the lapse of about
one quarter of a second are either discarded (if they
are not interesting) or passed on to the
short-term memory. This is the \textquotedblleft{cocktail effect}":
at a party we can pay attention to one
conversation, unaware of other conversations
until our name (or some gossip of interest) is
mentioned.

The short-term memory can store a limited
amount of information: about seven random letters,
or five random words, or four random phrases,
or one low resolution image. After up to eight to
twenty seconds this information is discarded
(forgotten) unless it is of our interest and
we refresh the memory by repetition (as when we
are given a new telephone number).

We can transfer information from the short-term
memory to the long-term memory. This transfer
requires that the information be classified and
related to other information in the long-term
memory. This is the learning process. As more
relations are established the easier it becomes
to find and recover information from the long-term
memory and \textquotedblleft{re-live}" it in the short-term memory.
The long-term memory can store practically an
unlimited amount of information which lasts for a
life time (it seems that the information is
lost only by disease or death).

We are aware of, or can pay attention to,
a limited amount of information. Therefore
we associate conscience with the short-term
memory.
Conscious perception takes hundreds of
milliseconds.

\textquotedblleft{Neurobiologists believe that memories are
encoded in the synapses, and two billion
synapses per square millimeter of cortex
can hold a lot of memories.}"\cite{Koch}

Most brain processes are unconscious. We have
perception and knowledge with and without awareness.

\textquotedblleft{Long-term memory enriches our lives
incredibly, but you don't need it to be aware.
All that's necessary for base-level awareness
is short-term memory and attention.}"\cite{Koch}

\textquotedblleft{When you superimpose a PET image on an MRI
image, you can see which brain structures are
active in a particular task.}"\cite{Koch} With this
technique it is discovered that different
brain areas are active when, for example,
we hear a word, see the word, speak the word
or write the word.\cite{Koch}

\begin{figure}
\begin{center}
\scalebox{0.6}
{\includegraphics{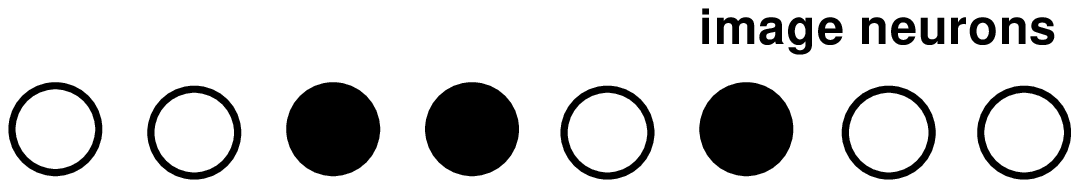}}
\caption{The filled circles represent image
neurons all over the cerebral cortex
that fire in response to a particular event
observed by the senses. The empty circles are image neurons
that do \textbf{not} fire in response to that particular event.}
\label{image_neuron}
\end{center}
\end{figure}

Let us quote Christof Koch again:
\textquotedblleft{So every time I see
an event, that event
corresponds to electrical activity all over the
brain. If I look at my friend Bill, say, his
face is represented in the brain area where
my face neurons are located, the hue of his
face is processed in V4, the fact that he's
moving around is represented in MT, my
memories of him correspond to activity in
the temporal lobes, and if he talks, his
speech activates my auditory cortex.}"
\textquotedblleft{Yet if I look at Bill I see a coherent whole.}"
\textquotedblleft{How come I don't get Bill's voice coming
from the man behind him?}"\cite{Koch}

Different qualities (form, movement, color, depth information from
binocular vision,
sound, smell, ...) are processed in parallel and
in hierarchical form in different parts of the brain.
Presumably at the highest level of these hierarchies
\textquotedblleft{all of the neurons responding to the stimulus
fire at roughly the same time.}"\cite{Koch}

Christof Koch and Francis Crick (yes, the same
Crick of the double helix) \textquotedblleft{think that if you are
aware of an event, \textit{all} the nerve cells involved
in the perception of that event anywhere in the
brain fire at the same time}"\cite{Koch} at about
40 pulses per second.
They also argue that  \textquotedblleft{to be aware of
something you need to attend to it, and you
need to put it into short-term memory}".\cite{Koch}

There does not appear to be a specific place in
the brain where consciousness resides. A model
of distributed processing appears to best
describe brain activity.

Conscience of the environment emerges in the
child as he or she acquires experience.
According to Piaget, before the first month of
age the reflexes lack will. From one to eight
months of age the child successively learns to repeat
voluntarily actions that cause him pleasure,
acquires the intention of his actions, searches
for hidden objects, and acquires long-term
goals. Self-consciousness emerges around the
eighteenth month of age when the child recognizes
himself in the mirror (seeing in the mirror that he
has a spot painted on the nose he reaches his nose).
At about the age of twenty four months the
child distinguishes the words \textquotedblleft{I}" and \textquotedblleft{you}".
So the brain is in constant development.

\begin{figure}
\begin{center}
\scalebox{0.6}
{\includegraphics{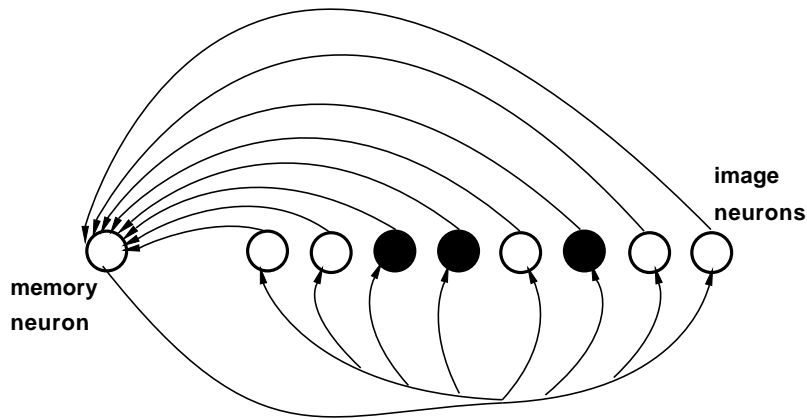}}
\caption{Image neurons, and one memory neuron before
being programed. The filled circles are image neurons that
fire in response to a particular event observed by the senses.}
\label{memory_neuron}
\end{center}
\end{figure}

\section{A model of memory and recognition}

The following is a simplified
model of memory and recognition that fits the
observations described in Section 2.

The model distinguishes two types of neurons
according to their different connections and
hence different functions. I will call these
neurons \textquotedblleft{image neurons}" and \textquotedblleft{memory neurons}".

When I see an event, each quality of the event
(shape, movement, color, distance, sound, smell...) is processed
in parallel, each in hierarchical form. At the
top level of these hierarchies are image
neurons.

When I see the event, image
neurons all over the cerebral cortex fire.
In Figure \ref{image_neuron} are shown three image neurons
that fire in response to a particular external stimulus.
The sub-set of image neurons that fires will be
called an \textquotedblleft{image}".

The crucial question is this one:
How can image neurons that fire all over the
cerebral cortex synchronize their pulses?

To answer this question we introduce a memory
neuron as shown in Figure \ref{memory_neuron}.
The unprogramed memory neuron has its dendrites
and its axon connected to many image neurons.

\begin{figure}
\begin{center}
\scalebox{0.6}
{\includegraphics{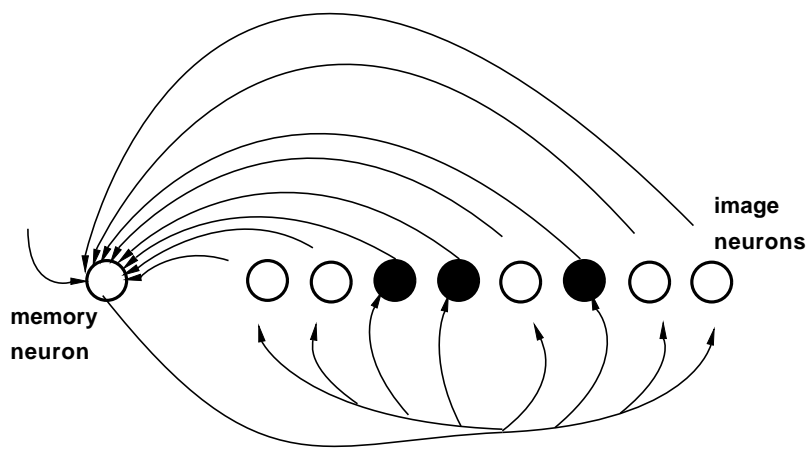}}
\caption{The same as Figure \ref{memory_neuron} but after the
memory neuron has been programed. Note that the synapses
connecting the memory neuron dendrites and axon to image
neurons that do \textbf{not} fire have been broken. The memory neuron
and the black image neurons form a positive feedback loop.
The black image neurons now fire in synchronism due to this
feedback loop.}
\label{programed_neuron}
\end{center}
\end{figure}

The memory neuron becomes programed and encodes
an image as follows.
The memory neuron breaks the synapses of its
dendrites and axon to the image neurons that are
\textbf{not} firing. See Figure \ref{programed_neuron}.
This breaking of
synapses is the storage of the image into
the long-term memory. Each memory neuron can
store one image. (It is important to mention that similar programing
of neurons has been studied in
kittens before the age of three months when the excess
of synapses are broken to obtain registered
binocular vision.\cite{Delbruck})
From the experiments we speculate that the programing of a memory
neuron takes of order $\approx1$ second.

The memory neuron serves three functions:
recognition, short-term memory, and
long-term memory as we now explain.

If I see the same event again, most of the same
image neurons fire. This will
cause the \textquotedblleft{memory neuron}" to \textquotedblleft{recognize}"
the image and fire. This is how we
recognize events.
The firing of the memory neuron in
turn \textbf{reinforces, synchronizes and completes} the firing
of the image neurons involved in the perception.
Thus a positive feedback loop
is established. Even when the stimulus is
removed the neurons keep firing in synchronism
(at about $40$ pulses per second\cite{Koch})
due to the positive feedback loop of the
memory neuron. The threshold of the memory
neuron raises (or, equivalently, the coupling weight of the synapses is
reduced) so that after up to eight to twenty
seconds the oscillation ceases. According to the model,
this is the mechanism of short-term memory.

Examples of the brain completing images are
the filling in of the blind spot, the
coloring of periferal vision and the
completion of a square.

If, for some reason, the memory neuron is
excited, then the image neurons involved in the
perception of the event fire in synchronism.
The memory neuron \textquotedblleft{recognizes}" this event,
fires, and establishes a positive feedback
loop. Self sustaining synchronized pulses occur
for a few seconds. This is how the event
is retrieved from the long-term memory
and is \textquotedblleft{re-created}"
or \textquotedblleft{re-lived}" in the short-term
memory. 

\begin{figure}
\begin{center}
\scalebox{0.6}
{\includegraphics{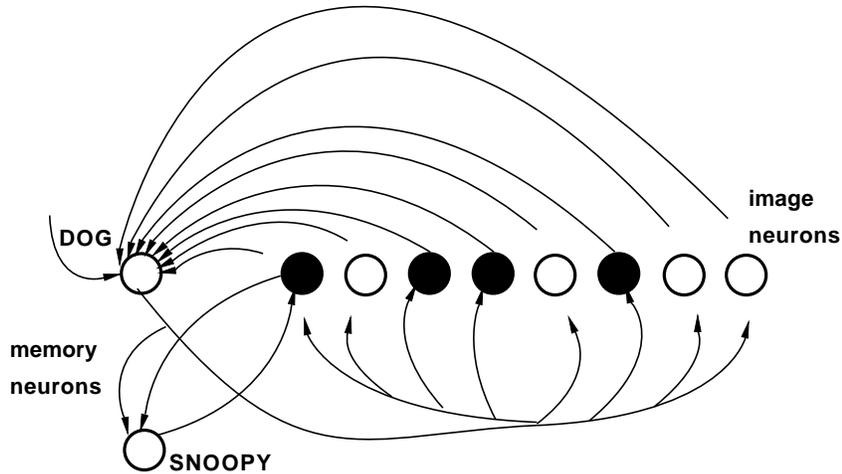}}
\caption{Two memory neurons are shown. The
memory neuron labeled \textquotedblleft{DOG}"
recognizes a dog in general. The memory neuron labeled
\textquotedblleft{SNOOPY}" recognizes a special dog.}
\label{hierarchy}
\end{center}
\end{figure}

There is a hierarchy of memory neurons,
so that if I see a dog coming towards me I
first recognize a dog in general, and then I recognize
the special dog Snoopy. The dog in general corresponds
to a particular set of image neurons firing in
synchronism. Snoopy corresponds to those same
image neurons plus additional image
neurons all firing in
synchronism. A simple hierarchy is shown in
Figure \ref{hierarchy}.

In Section 2 we mentioned that to transfer
information from the short-term
memory to the long-term memory it is necessary
to classify it and relate it to other information
already in the long-term memory. This classification
and relation is the construction of the
hierarchy of memory neurons such as the case
shown in Figure \ref{hierarchy}. According to the model,
this is the learning process.

We propose that there are memory neurons that recognize
parts of images, \textit{e.g.} a square, a face, the visual image
of a letter, the sound of a word, the smell of a rose...

Consider Figure \ref{image_neuron} again.
I see an external event through my senses.
As a result, image neurons
all over my cerebral cortex fire.
This firing is not in synchronism (yet). The qualities of the event
(shape, movement, color, sound...) have not been
related with each other. No memory neuron has recognized the image
yet.

After a few hundred milliseconds memory neurons \textbf{recognize} the image,
these memory neurons fire
and synchronize the pulses of the image neurons all over the
cerebral cortex. Positive feedback loops have been closed.
The qualities of the event have been \textbf{related with each other}.
The perception of the image no longer needs
input from the senses since the positive feedback is \textbf{self-sustaining}
for the next few seconds.
Because the memory neurons form part of a hierarchy, the information
of the image has been \textbf{related to other knowledge} in the
long-term memory. The image may now become conscious.

If the feedback loops are prevented from oscillating
we loose consciousness.
General anesthesia \textquotedblleft{shuts off}" consciousness by acting
on the synapses.
Likewise, the \textquotedblleft{center of sleep}", located
near the hypothalamus, regulates the
couplings at the synapses.

When a feedback loop oscillates we may become conscious of
the corresponding image.
The threshold of the memory neuron rises (or
equivalently the coupling at the synapses is
weekend) so that after a few seconds the oscillation
is no longer self-sustaining. Then another memory neuron
that codes a \textbf{related} image (and is therefore
close to threshold) can begin oscillating,
perhaps with the aid of input from the senses or noise.
Thus \textbf{one image
in conscience leads to a related image}. This is
the \textquotedblleft{stream of consciousness}".

\section{Estimates}

The following are rough order-of-magnitude
estimates. We humans have about $\approx10^{10}$ neurons.
Our eyes have $\approx10^8$ rods and cones. The retina,
which is part of the brain\cite{Feynman}, processes the visual
information so that \textquotedblleft{only}" $\approx10^6$ axons connect the
retina to the (rest of the) brain. Further processing will reduce
this number. So a rough estimate of the number
of image neurons (including all senses) is
of order $\approx10^5$. This number is of the same order
as the maximum number of dendrites of a neuron.
A very rough estimate of the number
of memory neurons is obtained as follows: assume
a memory neuron is programed, \textit{i.e.} an image is stored
into long-term memory, every $\approx10$ seconds (the duration
of short-term memory) for a lifetime.
The corresponding number of
memory neurons is of order $\approx10^8$.

Let us now consider a bee. It has of order
$\approx10^6$ neurons. The bee has eyes with a total
of $\approx27000$ facets. After processing, we will
need perhaps $\approx3000$ image neurons for all senses.
If $10\%$ of the bees neurons are memory neurons
then they can store up to $\approx10^5$ images in long-term
memory. So a worker bee that lives, say, six
months, can store into memory about one image
every minute or so. The bee will have some degree of consciousness.

\section{Conclusions}

The feedback loops shown in Figure \ref{hierarchy}
account for many observations on recognition, short-term
memory, long-term memory and the learning process, and
shed some light on consciousness.
We propose that some implementation of those feed back
loops is wide spread in the cerebral cortex.

We have not touched on the difficult questions: How does
feeling (such as pain) arise from the firing of neurons?
How do I become aware?
Can a silicon neural network feel pain and become aware of itself?
Does our attention
wander randomly from one image to a related one, or do we
have some control on our point of attention? How?

I hope the model serves as a point of departure for further
experimental and theoretical research on the brain.

\end{document}